\documentclass[useAMS,usenatbib]{mn2e}
\usepackage{times}
\usepackage{amssymb}
\newcommand{\simleq}{{\widetilde{>}}}

\newcommand{\kms}{km~s$^{-1}$}
\newcommand{\acc}{cm$^{-2}$}
\usepackage{graphics,epsfig}

\title[Faint dwarf galaxies]{Gas distribution, kinematics  and star formation in faint dwarf galaxies }
\author[
]
{
Ayesha Begum$^{1}$\thanks{E-mail:ayesha@ncra.tifr.res.in},
Jayaram. N. Chengalur$^{1}$,
I. D. Karachentsev$^{2}$,
S. S. Kaisin$^{2}$ 
\newauthor and M. E. Sharina$^{2}$
\\
\\
$^{1}$National Centre for Radio Astrophysics, Post Bag 3, Ganeshkhind, Pune 411 007, India\\
$^{2}$Special Astrophysical Observatory, Nizhnii Arkhys 369167, Russia\\
}
\begin{document}

\date{}


\maketitle


\begin{abstract}

        We compare the gas distribution, kinematics and the current star formation in
a sample of 10 very faint ($-13.37 < {\rm M_B} < -9.55$) dwarf galaxies. For 5 of these
galaxies we present fresh, high sensitivity, GMRT HI 21cm observations. We find that 
the large scale HI distribution in the galaxies is typically irregular and clumpy,
with the peak gas density rarely occurring at the geometric center. We also find
that the velocity fields for all the galaxies have an ordered component, although
in general, the patterns seen  do not fit that expected from a rotating disk.
For all our galaxies we construct maps of the HI column density at a constant
linear resolution of $\sim 300$~pc; this forms an excellent data set to check 
for the presence of a threshold column density for star formation. We find that
while current star formation (as traced by H$\alpha$ emission) is confined to
regions with relatively large (N$_{\rm HI} > (0.4 - 1.7)~\times~10^{21}$~\acc) HI column 
density, the morphology of the H$\alpha$ emission is in general not correlated
with that of the high HI column density gas. Thus, while high column density
gas may be necessary for star formation, in this sample at least, it is not
sufficient to ensure that star formation does in fact occur. We examine the
line profiles of the HI emission, but do not find a simple  relation between
regions with complex line profiles and those with on-going star formation. Our sample
includes examples of regions where there is on-going star formation, but the profiles
are well fit by a single Gaussian, as well as regions where there is no
star formation but the line profiles are complex. Finally, we examine the
very fine scale ($\sim 20 -100$~pc) distribution of the HI gas, and find that
at these scales the emission exhibits a variety of shell like, clumpy and filamentary 
features. The H$\alpha$ emission is sometimes associated with high density
HI clumps, sometimes the H$\alpha$ emission lies inside a high density shell,
and sometimes there is no correspondence between the H$\alpha$ emission
and the HI clumps. In summary, the interplay between star formation and
gas density in these galaxy does not seem to show the simple large scale 
patterns observed in brighter galaxies.

\end{abstract}


\section{Introduction}
\label{sec:intro} 

      In the currently popular hierarchical models of galaxy formation, star 
formation starts in small objects; these in turn later merge to form larger galaxies. 
In such a model, extremely small nearby galaxies are likely candidate ``primeval
galaxies'', in the sense that they may represent the earliest units of star formation 
in the universe. There is some observational support for these models, even in the 
very local universe, viz. (i)~the Milkyway itself appears to be still growing via 
the accretion of small companions like the Sagittarius dwarf galaxy (see e.g. Majewski et al.(2003)), 
and (ii)~nearby dwarf galaxies have stellar populations that are at least as old as 
the oldest stars in the Milkyway (see Grebel (2005) for a recent review).

      In detail, however, the star formation history of nearby dwarf galaxies appears 
to be extremely varied. At the two extreme ends, dwarf spheroidals have little 
gas or ongoing star formation while the relatively rare dwarf irregulars are gas rich 
and also generally have measurable ongoing star formation. Their past star formation
histories also appear to have been different -- at a given luminosity dwarf spheroidals 
are more metal rich than dwarf irregulars, indicative of rapid chemical enrichment 
in dwarf spheroidals in the past (Grebel (2004)). Why is it that dwarf irregulars, 
despite having a substantial reservoir of gas have resisted converting it into stars? 
What keeps the gas in dwarf irregulars from collapse? It is widely believed that the
smallest dwarf irregular galaxies have chaotic gas velocity fields (e.g. Lo et al.(1993)),
in this case the crucial question then becomes, what sustains these chaotic gas motions? 
In this context, it is interesting to note that for galaxies which have been observed 
with sufficient sensitivity and  velocity resolution, the velocity field has invariably
turned out to have a measurable ordered component, (Begum et al.(2003), Young et al.(2003), 
Begum \& Chengalur (2004)). 
Does this generally hold for extreme dwarf irregulars, or do some of them genuinely have 
no ordered components in their velocity fields? Irrespective of the exact nature of 
the velocity fields, the question  of why dwarf irregulars have been unable to 
convert their gas into stars remains. In spiral galaxies, the current star formation 
rate appears  to depend on at most two parameters (i)~the gas surface density and 
(ii)~some measure of the dynamical time. In practice, models which depend only on the 
gas surface density, such as the Schmidt star formation law, or those which depend 
on both these parameters, such as the Toomre's instability criterion (Toomre (1964)) 
appear to provide an equally good fit to the observations (Kennicutt (1998)). 
For irregular galaxies, Skillman (1987) has proposed that star formation occurs 
only above a threshold column density, and that this threshold may be related 
to a critical amount of dust shielding required for molecular gas formation. Are any 
of these models extrapolatable to the faintest dwarf irregulars? 

         We present here deep, high velocity resolution ($\sim$ 1.6 \kms) 
Giant Metrewave  Radio Telescope (GMRT) HI observations, as well as 
H$\alpha$ observations of a sample of faint (M$_{\rm B} \simleq-13.0$~mag)
galaxies, aimed at addressing the above issues. The rest of the paper is divided 
as follows. The dwarf galaxy sample is presented
in Sect.~\ref{sec:sample}, the GMRT  observations are detailed in Sect.~\ref{sec:obs}, 
while the results are presented in Sect.~\ref{sec:result} and discussed 
in Sect.~\ref{sec:discuss}.

\section{Dwarf galaxy sample}
\label{sec:sample}

\begin{table*}
\begin{center}
\caption{Optical parameters of the sample galaxies}
\label{tab:opt}
\vskip 0.1in
\begin{tabular}{|lcccccccccc|}
\hline
Galaxy&RA(J2000) & Dec(J2000)&M$\rm{_B}$ & D (Mpc) &B-V& R$\rm{_{Ho}}$ ($^\prime$) &i$\rm{_{opt}}$($^\circ$)&references\\
\hline
\hline
KK 44&04$^h$53$^m$06.9$^s$&  $+{67}^{\circ} 05' 57''$&$-$11.85&3.34&0.8& 0.7&65&1,3\\
KDG 52&08$^h$23$^m$56.0$^s$&  $+{71}^{\circ} 01' 46''$&$-$11.49&3.55&0.24& 0.65&24&3\\
UGC 4459&08$^h$34$^m$06.5$^s$&+${66}^{\circ} 10' 45''$&$-$13.37&3.56& 0.45&0.80&30&3,7\\
Leo A&09$^h$59$^m$26.4$^s$&  $+{30}^{\circ} 44' 47''$&$-$11.36&0.69&0.15& 3.5&54&3,4,8\\
CGCG 269-049&12$^h$15$^m$46.7$^s$&$+{52}^{\circ} 23' 15''$&$-$12.46&3.4&-&0.60&77&3\\
UGC 7298&12$^h$16$^m$28.6$^s$&$+{52}^{\circ} 13' 38''$&$-$12.27&4.21&0.29 &0.55&58&3,7\\
GR 8 &12$^h$58$^m$40.4$^s$&  $+{14}^{\circ} 13' 03''$&$-$12.11&2.10&0.32& 0.95&25&2,3\\
KK 230&14$^h$07$^m$10.7$^s$&+${35}^{\circ} 03' 37''$&$-$9.55&1.9&0.40&0.87&35&3\\
Sag DIG&19$^h$29$^m$59.0$^s$&  $-{17}^{\circ} 40' 41''$&$-$11.49&1.1&0.3& 1.8&45&3,6\\
DDO 210&20$^h$46$^m$51.8$^s$&  $-{12}^{\circ} 50' 53''$&$-$11.09&1.0&0.24& 1.8&62&3,5\\
\hline
\end{tabular}
\end{center}
{References:~
1-Begum et al. (2003),
2-de Vaucouleurs \& Moss (1983),
3-Karachentsev et al.(2004),
4- Karachentseva \& Sharina (1988)
5-Lee et al.(1999),
6- Lee \& Kim (2000),
7-Makarova (1999),
8-Tolstoy et al. (1998)
}
\end{table*}

The optical properties of our sample of ten galaxies are given in 
Table~\ref{tab:opt}. Fresh HI observations for five galaxies in the sample
viz. KDG 52, UGC 4459, CGCG 269-049, UGC 7298 and KK 230 are presented in this
paper. GMRT HI data for KK~44 (Camelopardalis~B), GR 8 and DDO 210 were
presented in our previous papers (Begum et al.(2003), Begum \& Chengalur 
(2003,2004)), although we include here fresh maps and measurements at 
angular scales that are relevant to the issues discussed in this paper. 
GR 8 and DDO 210 were also observed with the VLA; the VLA data for these
galaxies are presented in Young et al.(2003). VLA  HI data for Sag DIG and 
Leo~A were obtained from the VLA archive. These observations have been 
discussed earlier by Young \& Lo (1996) (Leo A) and Young et al.(1997) 
(Sag DIG); once again we present here only those maps and measurements that 
are relevant to this paper. 

\section{Observations and analysis}
\label{sec:obs}

\subsection{Optical observations and data analysis}
\label{ssec:halpha}

H$\alpha$ observations of some of our sample  galaxies, viz. KK~44, KDG~52, 
CGCG~269-049, UGC~7298 \& KK~230 were carried out at the 6-meter SAO
telescope using a 2048$\times$2048 pixel CCD camera. The scale was 
0.36 $\arcsec$/pixel, and the total area imaged was 6$\times6\arcmin$.
The H$\alpha$ + [NII] emission line fluxes were  obtained by
observing each galaxy through two filters: a narrow ($\sim$75 \AA)
interference filter centered on 6567 \AA, and a middle-width filter
($\lambda$ = 6063 \AA, $\Delta \lambda$ = 167 \AA) to determine
the nearby continuum level. The integration times were
2$\times300$ sec in the middle-width filter and 2$\times$600 sec in H$\alpha$.
Because the range of radial velocities was small, we used the same H$\alpha$ 
filter for all objects. The images were bias subtracted and flat fielded 
following standard procedures. After flat fielding, the next step was
to subtract the sky emission from both continuum and narrow band filter images. 
The continumm filter images were then scaled relative to the narrow band images 
using 5--10 unsaturated stars, and then subtracted from the narrow band 
filter images. The continuum-subtracted H$\alpha$ images were flux calibrated
using observations from the same night of two or more of Feige's photometric 
standards. Corrections for the Galactic extinction were made assuming
$\rm{A(H\alpha)}$= 2.32 E(B-V) using the data from Schlegel et al. (1998). 
The star formation rate for these galaxies  were calculated from the derived 
H$\alpha$ luminosities, using the conversion factor from Kennicutt (1998a)

\begin{equation}
\rm{SFR=7.9~\times~10^{-42}~L(H\alpha)~M_\odot yr^{-1}}
\end{equation}

The calculated SFR for our sample galaxies are given in Table~\ref{tab:result5}.
In case of  KDG 52 and KK 230, no H$\alpha$ emission was detected; the derived limits 
on the SFR for these galaxies is also listed in Table~\ref{tab:result5}.

   For Leo~A, Sag DIG and GR~8, H$\alpha$ images were downloaded from NED. Details of these 
images can be found in Hunter \& Elmegreen (2004). The H$\alpha$ image of UGC 4459 was kindly
provided by U. Hopp; details can be found in Schulte-Ladbeck \& Hopp (1998). For DDO~210, 
van Zee (2000)
detected a single source of H$\alpha$ emission in the galaxy; however follow up observations
suggested that it does not arise in a normal HII region, but probably comes from dense 
outflowing material from an evolved  star. In all the figures of DDO 210 in this paper, 
we show the location of this emission by a star, but caution the reader that it may not 
actually represent a star forming region.

Except for KK 230, broadband optical observations of all our sample galaxies were available
in the literature. For KK 230, V and I band HST ACS images were used to obtain the total
magnitude of the galaxy. The derived
magnitude is $\rm{I}(R < 40\arcsec)$ = 15$\fm$6$\pm0\fm15$, and the integrated  
$\rm{(V - I)}$ colour inside the same radius is $ 0.90$. Assuming a typical color $\rm{(B - V)} = 0.50$
for KK 230, we estimated its integrated blue magnitude to be $\rm{B} = 17\fm0\pm0\fm25$.

\subsection{HI observations and data analysis}
\label{ssec:obs}

\begin{table*}
\begin{center}
\caption{Parameters of the GMRT observations}
\label{tab:obs}
\vskip 0.1in
\begin{tabular}{|l|ccccccc|}
\hline
Galaxy&Date of&Velocity &Time on&synthesised & synthesised & Noise &Continuum Noise ($3\sigma$) \\
&observations&coverage&source&beam&  beam& &(26$^{\prime\prime} \times 22^{\prime\prime}$),($3^{\prime\prime} \times 3^{\prime\prime}$)\\
&&(\kms)&(hours)&(arcsec$^2$)& (pc$^2$)& (mJy)&(mJy) \\
\hline
\hline
KDG 52& 21-23, 27, &10 $-$ 220& 18&42$^{\prime\prime} \times39^{\prime\prime}$,26$^{\prime\prime}\times 23^{\prime\prime}$&723$\times$671, 447$\times$396& 1.7,1.5& 0.9, 0.42\\
& Jun 2002 && &16$^{\prime\prime}\times 15^{\prime\prime}$, 6$^{\prime\prime}\times 6^{\prime\prime}$&275$\times$258, 103$\times$103&1.3,0.9\\
UGC 4459& 15, 23, 24, &$-60~-~130$ & 14&45$^{\prime\prime}\times38^{\prime\prime}$,29 $^{\prime\prime}\times27^{\prime\prime}$&777$\times$656,500$\times$466&1.9,1.6 &1.0, 0.45\\
&Nov 2002 && &18 $^{\prime\prime}\times16^{\prime\prime}$,3 $^{\prime\prime}\times3^{\prime\prime}$&310$\times$276,52$\times$52 &1.4,1.2\\
CGCG 269-049& 23$-$25, &65 $-$ 275 & 16&42$^{\prime\prime} \times 39^{\prime\prime}$,28$^{\prime\prime} \times 24^{\prime\prime}$&692$\times$642, 461$\times$396 &2.0,1.8& 0.5, 0.3\\
& Nov 2002& & &18$^{\prime\prime} \times 17^{\prime\prime}$,$4^{\prime\prime} \times 3^{\prime\prime}$ &297 $\times$ 280, 66$\times$50&1.7,1.2\\
UGC 7298& 23$-$25,  &65 $-$ 275 & 16&42$^{\prime\prime} \times 37^{\prime\prime}$,26$^{\prime\prime} \times 24^{\prime\prime}$&857$\times$755, 530 $\times$490 &2.0,1.8& 0.5, 0.3\\
&Nov 2002 && &16$^{\prime\prime} \times 15^{\prime\prime}$,4$^{\prime\prime} \times 4^{\prime\prime}$&326 $\times$306, 82$\times$82&1.6,1.1\\
KK 230& 6 Jun, 8 May, &$-40 -$ 170 & 18&48$^{\prime\prime}\times45^{\prime\prime}$,34$^{\prime\prime}\times31^{\prime\prime}$&442$\times$ 415,313$\times$286 &1.6,1.4& 0.4, 0.2\\
&26 Nov  2001 && &26$^{\prime\prime}\times24^{\prime\prime}$,4$^{\prime\prime}\times3^{\prime\prime}$&240$\times$221, 37$\times$28 &1.2,0.8\\
\hline
\end{tabular}
\end{center}
\end{table*}

HI 21cm observations of KDG~52, UGC~4459, CGCG~269-049, UGC~7298 and KK~230  were conducted with the 
GMRT (Swarup et al. (1991)) between Nov.~2001 and Nov.~2002. KK~44, GR~8 and DDO~210  were also observed 
with the GMRT; details can be found in Begum et al.(2003), Begum \& Chengalur (2003,2004). 
Data for Sag DIG and Leo A were
obtained from the VLA archive. These observations are also discussed in Young \& Lo (1996,1997). Here
we briefly describe only the fresh GMRT observations.

    For all galaxies, the observing bandwidth of 1~MHz was divided into  128~spectral channels, yielding 
a spectral resolution of 7.81~kHz (velocity resolution of 1.65~\kms). The setup for the observations
is given in Table~\ref{tab:obs}. The flux and bandpass calibration were done using  3C48, 3C147 and 
3C286. The phase calibration was done once in every 30 min by observing the VLA calibrator sources
0831+557 (UGC~4459), 1216+487 (UGC 7298), 1216+487 (CGCG~269-049), 3C286 (KK 230) and 0834+555 (KDG~52).
The galaxies UGC~7298 and CGCG 269-049 are close in space ($\sim 12^\prime$) as well as in 
velocity, hence both were included in a single GMRT pointing (the field of view of the 
GMRT $\sim 24^\prime$).

\begin{figure*}
\psfig{file=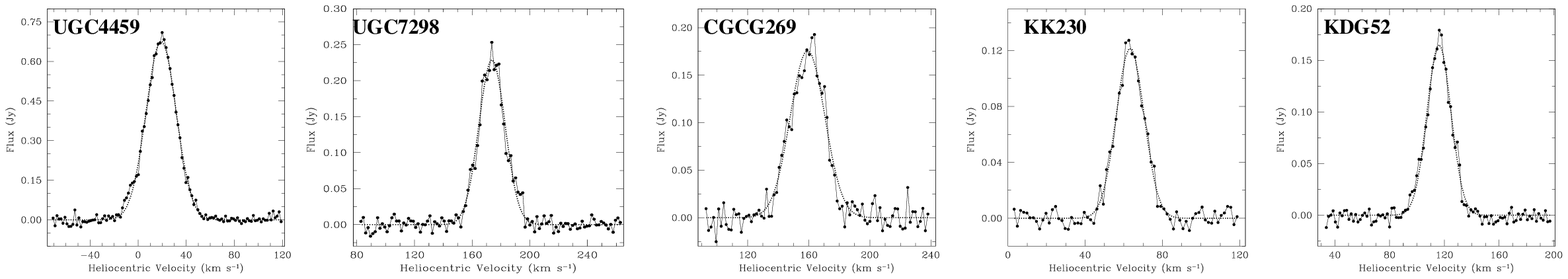,width=7.0truein}
\caption{The HI global profile for our sample galaxies derived from our coarsest resolution
HI distribution. The channel separation is 1.65 \kms. The dotted line shows a Gaussian fit to
the line profiles.}
\label{fig:spectra}
\end{figure*}

     The data were reduced in the usual way using standard tasks in classic AIPS.
For each run, bad visibility points were edited out, after which the data were
calibrated. The GMRT does not do online doppler tracking --  any required doppler
shifts have to be applied during the offline analysis. For UGC~7298 and CGCG~269-049,
the differential doppler shift over our observing interval was much less than
the channel width, hence, there was no need to apply any offline correction. On the
other hand, the differential shifts for UGC~4459, KK~230 and  KDG~52 were significant,
hence, for each of these galaxies, the calibrated (u,v) data set for each day was shifted in the
frequency space to the heliocentric velocity of the galaxy, using the task CVEL in AIPS.
For each galaxy, data for all the runs were then combined using the AIPS task DBCON.

The GMRT  has a hybrid configuration (Swarup et al. (1991)) which simultaneously 
provides both high angular resolution ($\sim 3^{''}$, if one uses baselines between 
the arm antennas) as well as sensitivity to extended emission (from baselines between 
the antennas in the central array). Data cubes were therefore made using various 
(u,v) cutoffs to get the images of HI emission at various spatial 
resolutions (see Table~\ref{tab:obs} for details). Except for the highest resolution HI 
data cubes for each galaxy, all the data cubes were deconvolved  using the AIPS 
task IMAGR. For the highest resolution data cubes in each galaxy, the signal to 
noise ratio (SNR) was too low for CLEAN to work reliably. Despite this, the low 
SNR of the images implies that the inability to deconvolve 
does not greatly degrade the dynamic range or fidelity of these images. The 
morphology of the emission in these galaxies should hence be accurately traced, 
apart from an uncertainty in the scaling factor (this essentially arises because 
the main effect of deconvolving weak emission at about the noise level corresponds 
to multiplying by a scale factor; see e.g. J\"ors\"ater \& van Moorsel (1995), Rupen (1999)).

     Continuum images were also made for all the galaxies by averaging the 
line free channels. No extended  ($26^{''}\times22^{''}$) or compact ($3^{''}\times3^{''}$)
emission  was detected from any of the galaxies. The $3\sigma$ limits for each galaxy are given 
in Table~\ref{tab:obs}.

       Moment maps  were made from the data cubes using  the AIPS task MOMNT. Maps of the velocity
field and velocity dispersion were also made in GIPSY using  single Gaussian
fits to the individual profiles. The velocity field produced by Gaussian
fitting is in reasonable agreement with that obtained from moment analysis.
The velocity dispersion ($\rm{\sigma_{obs}}$), as estimated by fitting single 
Gaussian component to the line  profiles  is given in Table~\ref{tab:result3}. 
In all cases, no  measurable variation of velocity dispersion was seen 
(within the errorbars) across each galaxy. This  lack of substantial variation 
of $\sigma$  across each galaxy is typical of such faint dwarf irregular galaxies 
(e.g. Begum \& Chengalur (2004), Begum et al. (2003), Skillman et al. (1988)).
As discussed in more detail in section~\ref{ssec:starformation}, single Gaussian
profiles are not necessarily a good fit throughout the galaxy; there
are regions where the emission profile is skewed or is otherwise
more complex than a single Gaussian.

\section{Results}
\label{sec:result}

\subsection{Large scale HI distribution and kinematics}
\label{ssec:HIdistribution}

       The global HI profiles for our sample galaxies, obtained from the coarsest resolution
data cubes (see Table~\ref{tab:obs}) are shown in Fig.~\ref{fig:spectra}.
Column (2$-$7) in Table~\ref{tab:result3} lists the parameters derived from the global HI profiles. 
Col.
(1)~gives the galaxy name, 
(2)~the integrated HI flux along with the errorbars, 
(3)~the velocity width at 50\% of the peak  ($\Delta V_{50}$), along with the errorbars, 
(4)~the central heliocentric velocity  (V$_\odot$) and its errorbars, 
(5)~ the HI mass along with its errorbars, 
(6)~ the HI mass-to-light ratio (M$\rm{_{HI}/L_B}$), 
(7)~the ratio of the GMRT flux to the single dish flux  (FI/FI$_{\rm{SD}}$). The single dish 
    fluxes for all the galaxies are taken from the tabulation in Karachentsev et al.(2004). 
    In the case of CGCG~269-049, single dish data is  not available. The parameters measured 
    from the GMRT HI profiles are in good agreement with those values obtained from the 
    single dish observations, in particular the HI flux measured at the GMRT agrees with 
    the single dish flux for all the galaxies. This indicates that no flux  was missed 
    because of the missing short spacings in our interferometric observations. 
Col.(8)~ shows the velocity dispersion, along with error bars, as measured from a single Gaussian fit
    to the line profiles,
(9)~represents the HI radius at a column density of $10^{19}$ atoms cm$^{-2}$,
(10)~the inclination as measured from the HI moment~0 maps,
(11)~the ratio of the HI diameter to the Holmberg diameter. For all the galaxies the HI emission 
     extends to $\sim2-3$ times the optical diameter, a typical ratio for dwarf irregular galaxies.

\begin{table*}
\begin{center}
\caption{Results from GMRT observations}
\label{tab:result3}
\vskip 0.1in
\begin{tabular}{|lcccccccccc|}
\hline
Galaxy& FI & $\Delta \rm{V_{50}}$ & V$_\odot$ & M${\rm{_{HI}}}$& ${\rm{M_{HI}/L_B}}$& $\rm{FI/FI_{SD}}$& $\rm{\sigma_{obs}}$&R$\rm{_{HI}}$ & i$\rm{_{HI}}$ & R$\rm{_{HI}/R_{Ho}}$\\
& (Jy \kms)&  (\kms)& (\kms) & $(10^{6}\rm{M_\odot})$& & & (\kms)& ($^\prime$) & ($^\circ$) & \\
\hline
\hline
KK 44&4.6(0.4)&21.4(1.0)&77.5(1.0)&12.2(1.2)&1.4&1.02&7.3(0.8)&1.6 &65&2.3\\
Leo A  &42.0(4.0)&18.8(0.7)&21.7(0.7)&4.7(0.4)&1.02&0.88&9.5(1.3)&8.0&62&2.3\\
KDG 52&3.8(0.4)&20.6(1.7)&116.0(1.9)&10.8(1.1)&1.8&0.85&9.0(1.0)&1.8&23&2.7\\
UGC 4459&21.5(2.2)&29.6(1.8)&19.2(2.3)&64.2(6.5)&1.4&1.01&9.0(1.6)&2.2&31&2.8\\
CGCG 269-049&4.7(0.5)&26.6(2.2)&159.0(3.4)&12.7(1.3)&0.9&$-$&9.5(1.0)&1.3&43&2.3\\
UGC 7298&5.2(0.5)&21.4(1.7)&174.0 (2.0)&21.6(2.1)&1.7&1.06&8.5(1.3)&1.8&28&3.1\\
GR 8    &9.0(0.9)&26.0(1.2)&217.0(2.2)&10.38(1.0)&1.02&1.03&9.0(0.8)&2.1&28&2.3\\
KK 230&2.2(0.2)&17.0(2.0)&63.3(1.8)&1.9(0.2)&1.9&0.86&7.5(0.5)&1.5&51&3.3\\
Sag DIG &23.0(1.0)&19.4(0.8)&$-$78.5(1.0)&5.4(0.2)&1.02&0.92&7.5(1.7)&2.1&33&2.3\\
DDO 210&12.1(1.2)&19.1(1.0)&$-$139.5(2.0)&2.8(0.3)&1.00&1.05&6.5(1.0)&2.4&27&1.3\\
\hline
\end{tabular}
\end{center}
\end{table*}

    The integrated HI emission of our sample galaxies overlayed on the optical 
Digitized Sky Survey  (DSS) images are shown in Fig.\ref{fig:m81dwa}[A]-
\ref{fig:kk230}[A].  The HI distribution in CGCG~269-049  
and KK~230 are dominated by a single clump of high column density, while the HI in UGC~4459 
and UGC~7298 are concentrated in two high-column-density  regions, separated by a low-column 
density region in the center. In the case of KDG~52, the HI is distributed in a clumpy,
incomplete ring. 

\begin{figure}
\begin{center}
\psfig{file=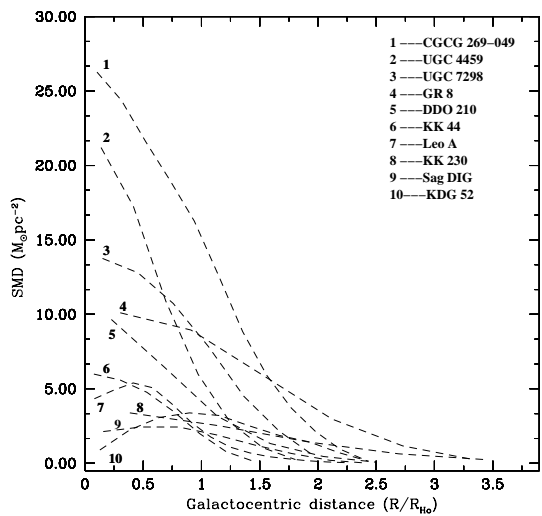,width=3.0truein}
\caption{The deprojected gas surface mass density (SMD) distribution for our sample 
galaxies. For all galaxies (except DDO~210, Sag DIG and Leo A), the SMD was computed for a linear resolution
of $\sim 500$~pc. For DDO~210, Sag DIG and Leo A the linear resolution is $\sim 300$~pc. The actual angular
resolutions are 29 $^{\prime\prime}\times27^{\prime\prime}$
(UGC 4459), 26$^{\prime\prime} \times 24^{\prime\prime}$ (UGC 7298), 28$^{\prime\prime} \times 
24^{\prime\prime}$ (CGCG 269-049), 48$^{\prime\prime}\times45^{\prime\prime}$ (KK 230), 
26$^{\prime\prime}\times 23^{\prime\prime}$ (KDG 52),41$^{\prime\prime}\times39^{\prime\prime}$ 
(GR 8), 61$^{\prime\prime}\times56^{\prime\prime}$ (DDO 210), 31$^{\prime\prime}\times
29^{\prime\prime}$ (KK 44), 67$^{\prime\prime}\times65^{\prime\prime}$ (Sag DIG) and 
78$^{\prime\prime}\times72^{\prime\prime}$ (Leo A). The gas SMD is obtained by scaling 
the HI SMD profile by 1.4 to account for primordial He.}
\label{fig:smd}
\end{center}
\end{figure}

   Inclinations (i$\rm{_{HI}}$) of our sample galaxies (except for KDG~52) were estimated from 
the HI moment~0 maps by  fitting elliptical annuli to the two lowest resolution images. For 
KDG~52, only the lowest resolution HI distribution is sufficiently smooth to be used for 
ellipse fitting. For all other galaxies, the inclination derived from these two resolution
images match within the errorbars. The estimated inclination for each galaxy (assuming  
an intrinsic thickness q$\rm{_{o}} =0.2$) is tabulated in Table~\ref{tab:result3}. 
Comparing this value to the optical inclination (Table~\ref{tab:opt}), shows that the two
inclinations are in agreement for UGC~4459 and KDG~52, whereas for the rest of the sample
galaxies the  optical inclination is either found to be much higher (UGC~7298 and 
CGCG~269-049) or lower (KK~230) than the inclinations derived from the HI morphology. 
Using the derived HI inclination, the deprojected HI radial surface mass density 
profiles (SMD) for each galaxy were obtained by averaging the HI distribution over 
elliptical annuli in the plane of the galaxy. The derived SMD profiles for each galaxy 
are given in Fig.~\ref{fig:smd}.

     We next discuss in detail the large scale HI distribution and kinematics for the
 five galaxies for which fresh GMRT observations are presented in the current paper.
 For similar details on the other galaxies in our sample, the reader is referred
 to Begum et al.(2003), Begum \& Chengalur (2003,2004) and Young \& Lo (1996,1997).

\subsection{Notes on individual galaxies}
\label{ssec:res1}

\subsubsection{KDG~52}
\label{sssec:m81dwa}

\begin{figure}
\begin{center}
\psfig{file=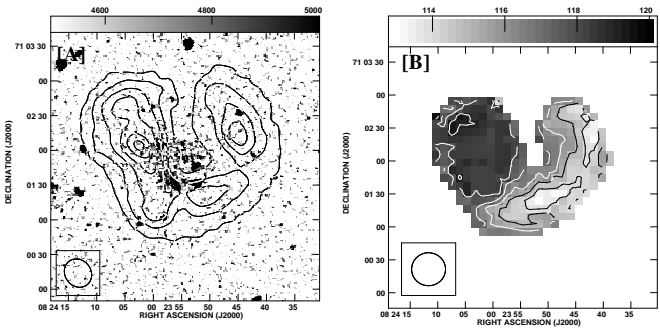,width=3.5truein}
\caption{{\bf{[A]}} The B band optical DSS image of KDG 52 (greyscales) with the GMRT
26$^{\prime\prime} \times 23^{\prime\prime}$ resolution integrated HI emission map
(contours) overlayed. The contour levels are 0.25, 1.00, 1.75, 2.49, 3.10 \& 3.67
$\times 10^{20}$ atoms cm$^{-2}$. {\bf{[B]}}  The velocity
field of the galaxy at $26^{''}\times23^{''}$ resolution. The contours are in the steps of
1.0 \kms and range from 113.0  \kms to 118.0 \kms.
}
\label{fig:m81dwa}
\end{center}
\end{figure}

KDG~52 (also called M81DwA) was discovered by Karachentseva(1968) and was later detected 
in HI by Lo \& Sargent (1979). 
The neutral hydrogen in this galaxy is distributed in a clumpy, broken ring surrounding
the optical emission (Fig.~\ref{fig:m81dwa}[A]). The central HI hole has a 
diameter of $\sim40^{\prime\prime}$ ($\sim$ 688 pc); similar central HI holes 
are seen in other faint dwarf galaxies (e.g. Sag DIG; Young \& Lo (1997), 
DDO~88; Simpson et al. (2005)). The HI hole is not exactly centered on
the optical emission;  the HI column density at the eastern side of the 
optical emission is   N$_{\rm HI} \sim 4 \times 10^{20}$ atoms cm$^{-2}$, while the rest of the 
optical emission lies inside the HI hole. Prior to this work, there have been 
two HI interferometric studies of KDG~52. It was observed with the WSRT 
by Sargent et al.(1983) with a velocity resolution of $\sim$ 8 \kms and later
re-observed with a high velocity resolution in the C array of  the VLA 
(Westpfahl et al. (1999)). The overall morphology of the earlier images 
compares well with that of our image.

Our coarsest resolution  HI distribution and velocity field (not shown)
shows faint emission in the center and in the northern region of the galaxy, a 
feature that is not visible at the higher resolutions. One may suspect that this 
HI emission is not real but is the result  of beam smearing. To check for this
possibility, the individual channel maps in the $42^{''}\times39^{''}$ data cube
were inspected. In the channel maps, the peak of the diffuse emission in the central
as well as in the northern region in the galaxy occurs at the same heliocentric velocity
as that of nearby HI clumps, suggesting that they may arise due to beam smearing. 
As a further check, the clean components from the $42^{''}\times39^{''}$ 
resolution data cube were convolved with a smaller restoring beam of $30^{''}\times30^{''}$, 
to generate a new data cube. The diffuse emission which was visible in $42^{''}\times39^{''}$ 
data cube is not seen in the channel maps in this cube, i.e. no clean components
were found in the region of diffuse emission. Finally, the HI flux 
measured from a genuine $30^{''}\times30^{''}$  resolution data cube (i.e. made from
the visibility data by applying the appropriate UV range and taper) is the same as that
measured from the $42^{''}\times39^{''}$ data cube. All these indicate that the diffuse 
emission in $42^{''}\times39^{''}$ is  entirely due to beam smearing. 

     The velocity field obtained from $26^{''}\times23^{''}$ resolution data cube
is given in Fig.~\ref{fig:m81dwa}[B]. The velocity field shows a large scale gradient 
across the galaxy with a magnitude of $\sim$ 1.7 \kms kpc$^{-1}$. However the velocity field
is clearly not consistent with pure rotation. One can still crudely
estimate the maximum possible circular velocity in the following way; the velocity 
difference from one edge of the galaxy to the other is $\sim 6$~\kms, this implies
that the magnitude of any circular velocity component must be limited to
V$\rm{_{rot} sin}(\it {i})\leq$~3  \kms.  Puche \& Westpfahl (1994) have tried
to model this velocity field, and find that a combination of rotation (with a 
magnitude of $7$~\kms) and expansion (with a magnitude of $5$~\kms) provides a 
reasonable fit. A similar combination of rotation and expansion was found to 
provide a good fit to the kinematics of another of our sample galaxies, viz.
GR 8 (Begum \& Chengalur (2003)).

KDG~52  is a member of M81 group of galaxies.  Bureau et al.(2004), have suggested that 
this galaxy  is probably  a tidal dwarf, 
formed through gravitational collapse of the tidal debris from the previous 
interactions of Holmberg~II with UGC~4483. In Sec.~\ref{ssec:HIkinematics},
we estimate the dynamical mass of this galaxy from the virial theorem; this
mass estimate implies that the galaxy has a significant amount of dark
matter. This would argue against a tidal dwarf origin for KDG~52, since
tidal dwarfs are generally not expected to be dark matter dominated (e.g. Braine et al.(2002)).

\subsubsection{UGC 4459}
\label{sssec:u4459}

\begin{figure}
\begin{center}
\psfig{file=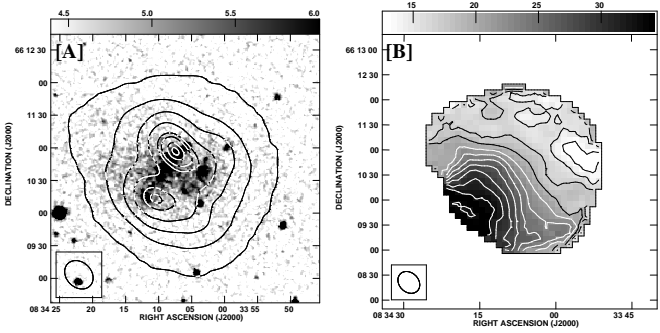,width=3.5truein}
\caption{{\bf{[A]}}
The B band optical DSS image of UGC 4459 (greyscales) with the GMRT  29$^{\prime\prime} 
\times 27^{\prime\prime}$ resolution integrated HI emission map (contours) overlayed. The contour
levels are 0.22, 3.33, 6.45, 9.56, 12.67, 15.79, 18.91, 22.00, 24.12, 28.25 and 31.36
$\times 10^{20}$ atoms cm$^{-2}$. {\bf{[B]}} The HI velocity field for galaxy at 
29$^{\prime\prime} \times 27^{\prime\prime}$ resolution. The contours are in the steps of 
2.0 \kms and range from 13.0  \kms to 33.0 \kms.
}
\label{fig:u4459}
\end{center}
\end{figure}

UGC~4459 is a member of M81 group of galaxies. It is relatively metal poor, 
with 12+log(O/H)$\sim7.62$ (Kunth \& O\"stlin (2000)). 
      The optical appearance of UGC~4459 is dominated by bright blue clumps, which emit
copious amounts of  H$\alpha$ (Fig.~\ref{fig:u4459}[A],~\ref{fig:overlay_HII_300pc}~\&
\ref{fig:overlay_HII}). The two high density 
peaks seen in the integrated HI map coincide with these star forming regions 
(Fig.~\ref{fig:u4459}[A]).

The velocity field of UGC~4459 (Fig.~\ref{fig:u4459}[B]) shows a large scale 
gradient (aligned along the line connecting the two star forming regions) across 
the galaxy. The magnitude of the average velocity gradient across the whole HI disk is
$\sim$ 4.5 \kms kpc$^{-1}$.  However we note that the gradient is not uniform across the 
galaxy. The receding (southeastern) half
of the galaxy shows a rapid change in velocity with galacto-centric distance, while
the approaching (northwestern) half of the galaxy shows a much more gentle gradient.
UGC~4459 is a fairly isolated dwarf galaxy with its nearest neighbor UGC~4483 at a 
projected distance of 3.6$^\circ$($\sim$ 223 kpc) and at a velocity difference of 
135~\kms. Being a member of the  M81 group, it is possible that interaction with 
intra-group gas could produce such disturbed kinematics. To check for this possibility,
we estimated the ram pressure required to strip gas from this galaxy. The threshold 
condition for ram pressure stripping is given by (Gunn \& Gott (1972))
\begin{equation}
\rho_{\rm{IGM}} v^2 \geq 2 \pi G \Sigma_{\rm{*}} \rm{\Sigma_{g}}
\label{eqn:rampressure}
\end{equation}
where, $\rho_{\rm{IGM}}$ is the density of the intra-group medium (IGM)
and $v$ is the relative velocity of the galaxy moving through the IGM. $\Sigma_*$
and $\rm{\Sigma_{g}}$ are stellar and  gas surface density respectively.
Taking $v\sim$ 190 \kms, typical for M81 group (Bureau \& Carignan (2002)),
and values for $\Sigma_*$ and $\rm{\Sigma_{g}}$ from the location in the galaxy
where the velocity field begins to look perturbed, we find that the IGM volume 
density required to strip the ISM from UGC 4459 is n$_{\rm{IGM}}\geq 8 \times 
10^{-5}$ cm$^{-3}$. UGC~4459 is located at a projected separation of 
$\sim8.4^\circ$ (520~kpc) to the South-West of M81 (which we can take to be the
center of the M81 group). The n$_{\rm{IGM}}$ required for ram pressure stripping of 
UGC~4459 is much higher than n$_{\rm{IGM}}$ expected at this location 
($\sim1.4\times10^{-6}$ cm$^{-3}$; assuming that 1\% of the virial 
mass of the group is dispersed uniformly in a hot IGM within a sphere just enclosing UGC 4459; 
Bureau \& Carignan (2002)). Hence, it seems unlikely 
that the peculiar kinematics of the galaxy is due to IGM ram pressure.

   Given the kinematical asymmetry between the two halves of the galaxy, it is not
surprising that a tilted ring fit does not give consistent results for the two
halves. The difference in the peak velocities for the rotation curves derived 
from the two halves is $\sim 15$~\kms. This difference is  significant compared to the peak 
value of $25$~\kms obtained for the receding half of the galaxy. One can crudely
estimate the maximum possible circular velocity in the following way; the velocity 
difference from one edge of the galaxy to the other is $18$~\kms, this implies
that the magnitude of any circular velocity component must be limited to
V$\rm{_{rot} sin}(\it {i})\leq$~9  \kms. 

    Pustilnik et al.(2003) found substantial small scale velocity gradients in the H$\alpha$ 
emission along a slit placed parallel to the optical major axis (i.e. also along 
the direction of maximum velocity gradient in the HI velocity field), as well
as a large scale gradient, with magnitude somewhat larger than what we
observe in the HI.

UGC~4459 has the largest star formation rate of all the galaxies in our sample.
Pustilnik et al.(2003) estimate very young ages ($\sim 3 -8$~Myr) for the star 
forming knots in the galaxy. Since they find no nearby galaxy that could have
triggered this recent starburst, they suggest that it could be triggered by
tidal interaction with the M81 group as a whole, or by interaction with
the intra group medium. As we argued above, the ram pressure of the intra group
medium is likely to be small. The velocity field of UGC~4459 is however qualitatively
very similar to that of DDO~26 (Hunter \& Wilcots (2002)) and IC~2554 (Koribalski et al.(2003)), both
of which are suspected to be late stage mergers. It seems possible therefore
that UGC~4459 too represents a recent merger of two still fainter dwarfs.

\begin{figure}
\begin{center}
\psfig{file=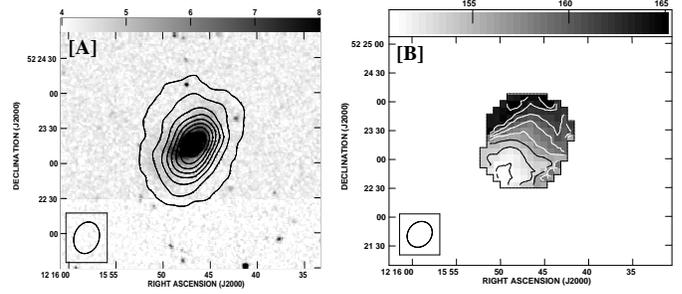,width=3.5truein}
\caption{{\bf{[A]}} The B band optical DSS image of CGCG 269-049 (greyscales) with the GMRT
28$^{\prime\prime} \times 24^{\prime\prime}$ resolution integrated HI emission map (contours)
overlayed. The contour levels are 0.08, 1.12, 2.18, 3.23, 4.28, 5.33, 6.39, 7.44, 8.49 and
9.55 $\times 10^{20}$ atoms cm$^{-2}$. {\bf{[B]}} The HI velocity field for galaxy at 
28$^{\prime\prime} \times 24^{\prime\prime}$ resolution. The contours are in the steps 
of 2.0 \kms and range from 151.0  \kms to 165.0 \kms.
}
\label{fig:cgcg}
\end{center}
\end{figure}

\subsubsection{CGCG 269-049}
\label{sssec:cgcg}

CGCG 269-049 is an extremely metal poor dwarf galaxy with 12+log(O/H)$\sim7.43$ 
(Kniazev et al. (2003)). It is a member of the Canes Venatici~I cloud.
The optical emission in CGCG 269-049 shows two components, a central
compact component and an outer faint extended component; both elongated
in the northwest direction. The HI distribution of the galaxy also
shows an elongation in the same direction. However, a misalignment 
of $\sim10^\circ$ is seen between the optical and the HI major axis. 
The HI distribution is regular and shows a slightly off-centered peak; 
this signature is more prominent in the high resolution HI images.

CGCG 269-049 is undergoing a burst of star formation as indicated by  
strong emission lines in its spectra. It has been suggested that
starbursts in dwarf galaxies could be triggered by tidal
interaction with a companion (Taylor (1997), Walter \& Brinks (2001)). 
While CGCG~269-049 does have a nearby companion (viz. UGC~7298 as discussed
in Sect.\ref{sssec:u7298}) the HI distribution in neither of these galaxies is suggestive
of tidal interaction.

The velocity field of the galaxy shows a large scale gradient, roughly
aligned with the morphological major axis and with a magnitude of 
$\sim$ 5.2 \kms kpc$^{-1}$ (Fig.~\ref{fig:cgcg}[B]). Of all the galaxies
in this subsample, CGCG~269-049 has a velocity field that most resembles
that expected from a rotating disk. Substantial deviations from simple
rotation can however be seen, and a tilted ring fit to the velocity
field does not yield meaningful results. CGCG~269-049 has one of the
highest current star formation rates among our sample galaxies; 
H$\alpha$ imaging of this galaxy shows a bright H$\alpha$ core near
its center (see Fig.~\ref{fig:overlay_HII_300pc} \& \ref{fig:overlay_HII}). 
From  Fig.~\ref{fig:cgcg}[B], one can see kinks in the velocity field in 
the regions near this star forming knot. Hence, it is likely that the 
energy input from  the on-going star formation  in the galaxy is responsible 
for at least some of the distortions seen in the velocity field.
The maximum velocity difference from one edge of the galaxy to 
the other is $\sim$ 16 \kms, hence  V$_{\rm{rot}}\rm{sin}{\it (i)}\leq8$ \kms.

\subsubsection{UGC 7298}
\label{sssec:u7298}

\begin{figure}
\begin{center}
\psfig{file=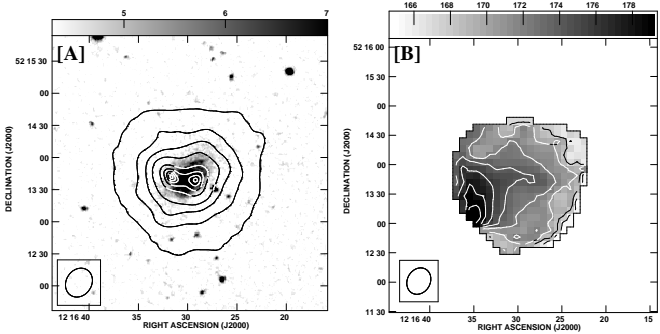,width=3.5truein}
\caption{{\bf{[A]}} The B band optical DSS image of UGC 7298 (greyscales) with the GMRT  26$^{\prime\prime} 
\times 24^{\prime\prime}$ resolution integrated HI emission map (contours) overlayed. The contour
levels are 0.2, 2.7, 5.3, 7.9, 9.1, 10.5, 11.2 and 11.5 $\times 10^{20}$ atoms cm$^{-2}$.
{\bf{[B]}} The HI velocity field for galaxy at 26$^{\prime\prime} \times 24^{\prime\prime}$ resolution.
The contours are in the steps of 2.0 \kms and range from 166.0  \kms to 180.0 \kms.
}
\label{fig:u7298}
\end{center}
\end{figure}

\begin{figure*}
\psfig{file=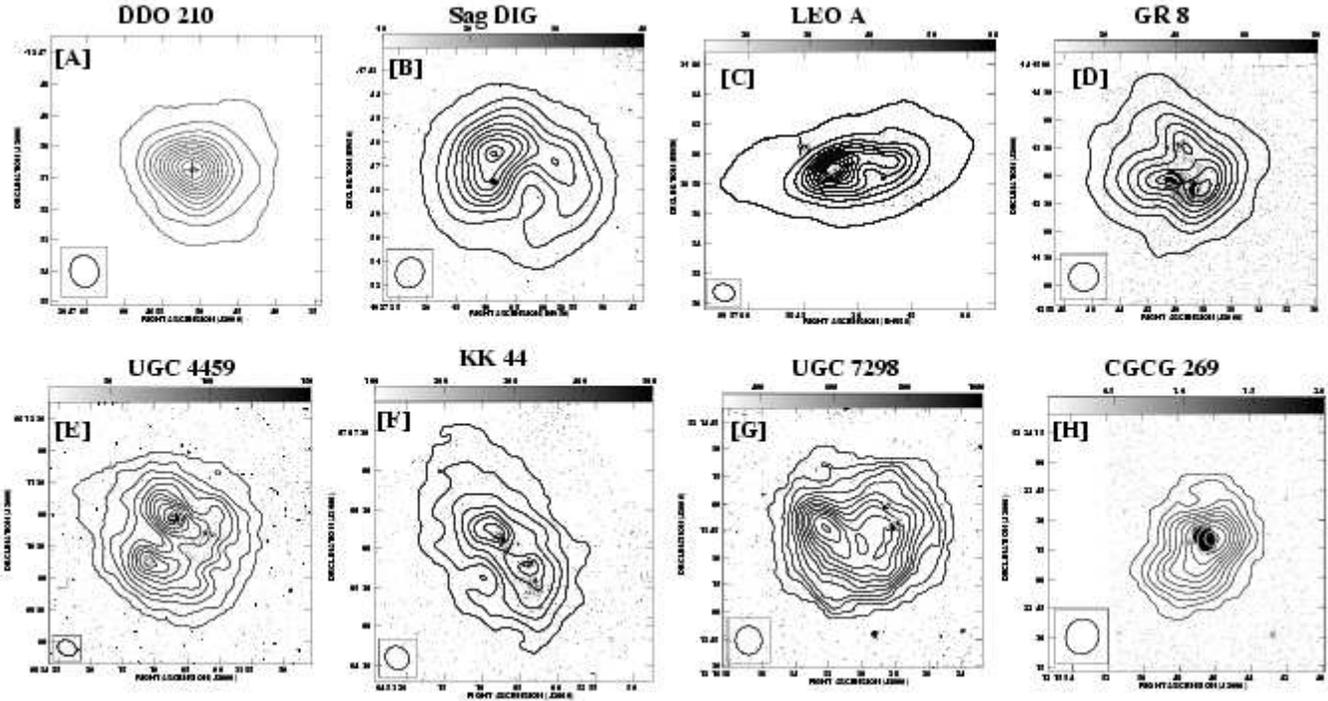,width=7.0truein}
\caption{ Comparison of integrated HI emission (contours)  from the sample galaxies at a 
linear resolution of $\sim$300 pc with  H$\alpha$ emission (greyscales). The angular resolution of 
the HI images are given in Table~\ref{tab:result5}. KK 230 and KDG 52 are not shown in the figure 
as no H$\alpha$ was detected in these galaxies. In the case of DDO 210, the region of H$\alpha$
emission found in the galaxy (van Zee 2000) is show as a star, however we note that this may not
be an HII region. The H$\alpha$ image of UGC 4459 is kindly provided by  U. Hopp and is previously 
published by Schulte-Ladbeck \& Hopp (1998). H$\alpha$ images of 
Leo A, Sag DIG and GR 8 are obtained from NED. The observations of these galaxies is described
in Hunter \& Elemgreen (2004). For UGC 7298, CGCG 269-049 and KK 44 the alignment of the H$\alpha$ 
images and radio images is done by comparing the continuum+H$\alpha$ images with the optical DSS 
images. In case of Leo A and Sag DIG in the absence of continuum images,
alignment of the H$\alpha$ images and radio images is done using the coordinates of the HII regions from 
Strobel et al.(1991), whereas for GR 8 the coordinates of the HII regions were obtained from Hodge et al.
(1989)}
\label{fig:overlay_HII_300pc}
\end{figure*}

UGC 7298 is a member of the Canes Venatici~I  cloud of galaxies. 
The velocity field of UGC~7298  (Fig.~\ref{fig:u7298}[B]) shows a large scale gradient 
roughly aligned with the line joining the two high density gas clumps. The magnitude
of the gradient is $\sim$ 3.5 \kms kpc$^{-1}$. The velocity field is broadly
similar to that in UGC~4459. The optical properties of the two galaxies are however
very dissimilar. UGC~4459 is currently undergoing a starburst and its optical
appearance is dominated by bright star forming knots. UGC~7298 on the other hand
has a very small current star formation rate, as inferred from  very faint H$\alpha$ 
emission in the galaxy (Fig.~\ref{fig:overlay_HII_300pc}~\& \ref{fig:overlay_HII}).
The maximum velocity difference from one edge of the  galaxy to the other is $\sim$ 
16 \kms, implying that  V$_{\rm{rot}} \rm{sin}{\it (i)}\leq8$ \kms. 

\subsubsection{KK~230}
\label{sssec:kk230}

\begin{figure}
\begin{center}
\psfig{file=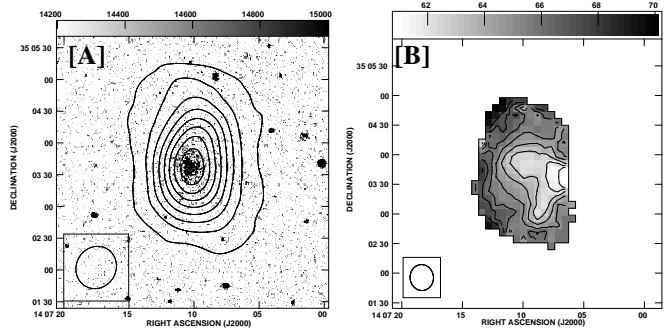,width=3.5truein}
\caption{{\bf{[A]}} The B band optical DSS image of KK 230 (greyscales) with the GMRT
48$^{\prime\prime} \times 45^{\prime\prime}$ resolution integrated HI emission map (contours)
overlayed. The contour levels are 0.09, 0.50, 0.92, 1.33, 1.75, 2.16, 2.58, 3.00, 3.41 and 3.83 $\times 
10^{20}$ atoms cm$^{-2}$. {\bf{[B]}} The HI velocity field for galaxy at 26$^{\prime\prime} 
\times 24^{\prime\prime}$ resolution.
The contours are in the steps of 1.0 \kms and range from 61.0  \kms to 68.0 \kms.
}
\label{fig:kk230}
\end{center}
\end{figure}

KK~230, the faintest dwarf irregular galaxy in our sample, is yet another
member of the Canes~Venatici~I cloud of galaxies (Karachentsev et al. (2003)). 
The velocity field (Fig.~\ref{fig:kk230}[B])
shows a gradient  in the east-west direction (i.e. roughly perpendicular to the HI and optical 
major axis) with a magnitude of $\sim 6$~\kms~kpc$^{-1}$. Even apart from this misalignment,
the velocity field bears little similarity from that expected from a rotating axi-symmetric
disk. The origin of the velocity gradient in KK~230 is rather puzzling. This galaxy has
no measurable ongoing star formation, and no H$\alpha$ emission was detected from the galaxy.
It also lies at the periphery of the Canes Venatici~I cloud group of galaxies;
Karachentsev et al. (2004) found its  tidal index to be $-$1.0, meaning that it is
a fairly isolated galaxy. This, along with the fairly regular HI distribution 
make it unlikely that tidal forces are responsible for the observed velocity
field. The maximum velocity difference from edge of galaxy to the other is 
$\sim 10$~\kms, which gives  V$\rm{_{rot} sin}(\it {i})\leq$~5  \kms.

\section{Discussion}
\label{sec:discuss}

\subsection{Dynamical mass of our sample galaxies}
\label{ssec:HIkinematics}

         As discussed above, large scale systematic gradients are seen across 
all the newly mapped galaxies. In fact,  all the 10 galaxies in our sample have
velocity fields with  a measurable ordered component, contrary to the general 
belief (e.g. Lo et al. (1993)) that the velocity fields of faint dwarf galaxies 
are  ``chaotic". Some of our sample galaxies overlap with those in Lo et al. (1993), 
and from a comparison of the new and old determinations of the velocity fields, 
it appears that high sensitivity and high velocity resolution ($\sim 1.6$~\kms as
opposed to the earlier used $\sim$ 6~\kms) are crucial to discern systematic 
kinematical patterns in such faint galaxies. 

         The origin of these ordered fields is unclear. One would expect that
in the absence of external forces, or internal energy input, gas with non zero
angular momentum would settle down into a rotating disk. Tidal forces, and/or 
energy input from star formation could profoundly disrupt the gas velocity fields. Indeed
it has often been suggested that a strong starburst could drive out the ISM
of such small galaxies (see e.g. Dekel \& Silk (1986), Efstathiau (2000),
Ferrara \& Tolstoy 2000). For two of the galaxies in our sample, viz.
GR 8 (Begum \& Chengalur (2003)) and  KDG~52 (Puche \& Westpfahl (1994)), detailed modeling
shows that the velocity field can be fit by a combination of circular and
radial motions of the gas. In general though, there does not 
seem to be any particular correlation between the current star formation
rate (see Table~\ref{tab:result5}) and the distortion of the velocity fields.
For example, the velocity field of CGCG~269-049 (Fig.\ref{fig:cgcg}[B]) shows relatively
mild deviations from that expected from rotation, as compared to
that of UGC~7298 (Fig.~\ref{fig:u7298}[B]), even though both galaxies have 
comparable luminosities and the star formation rate in CGCG~269-049 is
more than an order of magnitude more than that of UGC~7298. Tidal
interactions are also not clearly implicated, as several of our galaxies
are relatively isolated, and none of them show morphologies typical
of tidal interactions. Interestingly, some compact high velocity clouds
(notably M~31~HVC~1, M~31~HVC~16; Westmeier et al.(2005)) show similar
velocity gradients. Westmeier et al.(2005) argue against these
velocity gradients being due to tidal forces, and suggest that they
may be indicative of dark matter in these objects.

           Given their peculiar kinematics, it is difficult to accurately 
determine the total dynamical mass of our sample galaxies. We instead 
compute an indicative dynamical mass using the virial theorem, assuming 
that the HI distribution is spherical and has an isotropic velocity 
dispersion and negligible rotation. We realize that these assumptions
are unlikely to be rigorously justifiable in the current situation,
and therefore this mass estimate is at best indicative. Under
the assumptions above, the virial mass estimate is (e.g. Hoffman et al. (1996)):

\begin{equation}
\rm{M_{VT}=  {5~R_{HI} \times \sigma_{\rm true}^2 \over  G}}
\label{eqn:virial}
\end{equation}

where $\rm{R_{HI}}$ is the HI radius of the galaxy at a column density of 10$^{19}$ atoms 
cm$^{-2}$ (from Table~\ref{tab:result3}). $\sigma_{\rm true}$ is the  HI velocity 
dispersion corrected for the
instrumental broadening as well as  for the  broadening due to the  velocity
gradient over the finite size of the beam. This correction is applied using

$\sigma_{\rm true}^{2}=\sigma_{\rm obs}^2-\Delta v^2-
        \frac{1}{2}{b}^2{({\nabla}v_{\rm{o}})}^2$,

where  $\sigma_{\rm true}$ is the true velocity dispersion, $\Delta v$ is the channel width,
$b$ characterizes the beam width (i.e. the beam is assumed to be of form $e^{-x^2/b^2}$)
and $v_{\rm{o}}$ is the observed rotation velocity. $\sigma_{\rm obs}$ is the observed 
velocity dispersion in the HI gas  given in Table~\ref{tab:result3}.

\begin{table}
\begin{center}
\caption{Dynamical mass estimate for our sample galaxies}
\label{tab:result4}
\vskip 0.1in
\begin{tabular}{|lccccc|}
\hline
Galaxy&$\sigma_{\rm true}$& $\Gamma_{\rm{B}}$ &M$_*$ &M$\rm{_{{VT}}}$\\
&(\kms)&& ($10^6 \rm{M}_\odot$)& ($10^8 \rm{M}_\odot$)\\
\hline
\hline
KDG 52       &8.5 &0.6 & 3.7 &1.5 \\
UGC 4459     &8.0 &1.0 &45.7 &1.9 \\
CGCG 269-049 &9.0 &1.0 &15.0 &1.2 \\
UGC 7298     &8.0 &0.7 & 8.8 &1.6 \\
KK 230       &7.0 &1.0 & 1.0 &0.5 \\
\hline
\end{tabular}
\end{center}
\end{table}

\begin{figure}
\begin{center}
\psfig{file=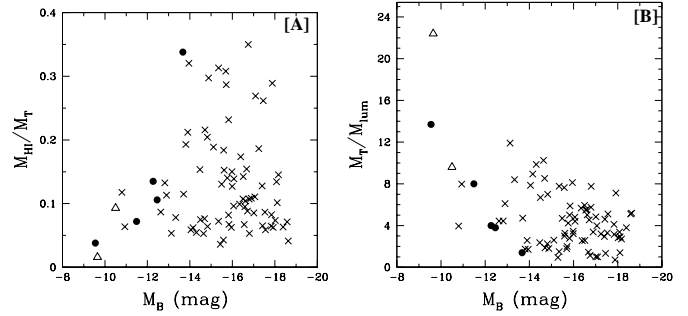,width=3.5truein}
\caption{
{\bf{[A]}} $\rm{M_{HI}/M_{T}}$ as a function of M$\rm{_{B}}$ for a sample of 
	   dwarf irregular galaxies ($\rm{M_B\geq -19}$). $\rm{M_{T}}$ is estimated from
	   the last measured point of the rotation curve for Verheijen(2001), Swaters 
	   (1999) and Cote et al. (2000) (represented as crosses), whereas for galaxies from 
	our sample (solid dots) and for LGS-3 and Sag DIG from Young \& Lo (1997) 
	(triangles), $\rm{M_{T}}$  is estimated from the virial theorem.    
{\bf{[B]}} $\rm{M_{T}/M_{lum}}$ for the same sample. See text for more details.
}
\label{fig:sample_corr}
\end{center}
\end{figure}

Table~\ref{tab:result4} lists the  estimated dynamical mass ($\rm{M_{VT}}$)
for our sample  galaxies. The columns in the table are:
Col.(1)~The galaxy name,
Col.(2)~The estimated $\sigma_{\rm true}$,
Col.(3)~The stellar mass-to-light ratio, $\rm{\Gamma_B}$ is obtained from the observed  
        B$-$V colour for each galaxy, using the low metallicity Bruzual \& Charlot SPS 
        model for a stellar population with metallicity Z=0.008, a Salpeter IMF  and an
        exponentially declining star formation rate of age 12 Gyrs (Bell \& de Jong 2001). 
        In absence of any colour information for CGCG~269-049, we assume $\rm{\Gamma_B}$=1 for it.
Col.(4)~The stellar mass obtained from the assumed mass-to-light ratio
and Col.(5)~The virial mass, as obtained from Eqn.~\ref{eqn:virial}.

Figure~\ref{fig:sample_corr}[A] is a plot of the ratio of HI mass to the dynamical mass 
against absolute blue magnitudes for a  sample of dwarf irregular galaxies. The 
references from which the data are taken are listed in the figure caption. For
our current sample, the dynamical masses are taken from Table~\ref{tab:result4},
Begum et al.(2003), Begum \& Chengalur (2004) (for KK~44 and DDO 210) and 
Young \& Lo (1997) (for LGS-3 and Sag DIG).
The ratio of M$\rm{_{HI}/M_{VT}}$ for UGC~4459 is found to be $\sim$0.34, which 
is larger than a value typically seen in dwarf galaxies. Such high value of 
M$\rm{_{HI}/M_{VT}}$ is also seen in  some blue compact dwarf galaxies 
(e.g. van Zee et al. (1998)). Figure~\ref{fig:sample_corr}[B] shows  M$\rm{_{VT}/M_{lum}}$ 
for the same sample, plotted as a function of M$\rm{_B}$. The luminous mass, 
$\rm{M_{lum}}$ is the sum of the stellar and gas mass. The stellar mass for
all galaxies was computed in exactly the same way as for our sample galaxies;
the gas mass is obtained  by taking into account the contribution of primordial 
He i.e. M$\rm{_{gas}=1.4\times M_{HI}}$.   We note that although M$\rm{_{VT}/M_{lum}}$ 
 does not show any  correlation with M$\rm{_B}$ for the full sample, there is a trend of an increase 
in the M$\rm{_{VT}/M_{lum}}$ ratio with the decrease in M$\rm{_B}$, seen in 
our sample galaxies. Further, a jump in M$\rm{_{VT}/M_{lum}}$  ratio is seen at 
the faintest luminosities  (M$\rm{_B}>-10.0$). While this might be indicative
of increased baryon loss from the  halos of these galaxies (e.g. Gnedin et al. (2002)), we 
caution that there is considerable uncertainty in the dynamical mass for
the galaxies in this magnitude range, and also that the total number of galaxies
is too  small to substantiate such a claim.

\subsection{HI column density and star formation}
\label{ssec:starform}

      Our sample galaxies have widely different star formation rates 
(see Table~\ref{tab:result5}), and range from having no detectable on-going 
star formation (e.g. KK~230, KDG~52), to having an optical appearance that 
is dominated by bright star forming knots (e.g. UGC~4459, CGCG~269-049). 
As such, this is a well suited sample for trying to determine the 
connections (if any!) between the HI distribution and kinematics 
and star formation.

            For spiral galaxies, the star formation rate appears to be quite
well correlated with the gas column density, though it is unclear if the
gas column density is the only relevant parameter, as in the Schmidt law,
or whether a combination of gas column density and the dynamical time are
important, (as in the Toomre's instability criteria, see Kennicutt (1998)).
For dwarf irregular galaxies, it has been suggested that star formation
occurs only above a critical threshold column density ($\sim 10^{21}$~\acc,
when measured at a spatial resolution of $\sim 500$~kpc) and that this
may be because a critical amount of dust shielding is required before
star formation can commence (Skillman (1987)). Since the observed 
column density is resolution dependent and that the distance of our sample 
galaxies varies from  0.70 Mpc ( Leo~A) to 4.2 Mpc (UGC~7298), one would
require maps at angular resolutions varying by a factor of $\sim 6$. 
This problem is well suited to a telescope like the GMRT, where because
of the hybrid configuration, maps at a range of angular resolutions can
be made from a single observing run. For all our galaxies,
we produced CLEANed HI maps at an angular resolution corresponding to a
linear scale of $\sim 300$~pc at the distance to the galaxy. We thus have 
a unique data set which spans a wide range of star formation rates, 
but for which maps are available at a similar linear resolution. The 
results from a comparison of the HI distribution  with the sites of H$\alpha$ emission 
are given in Table~\ref{tab:result5}.  
Col.(1)~shows the galaxy name, 
Col.2)~the absolute blue magnitude (M$\rm{_B}$) of the galaxy, 
Col.3)~the resolution of the HI column density map, 
Col.4)~the corresponding linear resolution in pc, 
Col.5)~the observed HI peak column density, 
Col.6)~the peak gas surface density. The gas surface densities are obtained by 
       correcting the observed HI column densities for inclination and for primordial
       He content (which we take to be  10\% of HI by number). Note that the peak
       gas density need not occur at the center of the galaxy, 
Col.7)~the column density of the HI contour that just encloses all the HII emission 
       in the galaxy,
Col.8)~ current star formation rate (for KDG 52 \& KK 230 the limits on the star 
       formation rate are listed), 
Col.9)~the metallicity of the galaxy (for some of our sample galaxies no 
      measurement of the metallicity exist)  and 
Col.(10)~references for the star formation rate and the metallicity. 

\begin{table*}
\begin{center}
\caption{Comparison of HI and optical emission from the sample galaxies}
\label{tab:result5}
\vskip 0.1in
\begin{tabular}{|lccccccccc|}
\hline
Galaxy& M$\rm{_B}$&Beam  & Beam& N$\rm{_{HI}^{Peak}}$ & $\Sigma\rm{_{g}^{peak}}$ &$\Sigma\rm{_{HI}^{\rm c}}$(HII)  & Log[SFR]&12+log(O/H)&References \\
& &(arcsec) & (pc)& (10$^{21}$ cm$^{-2}$)& (10$^{21}$ cm$^{-2}$)&(10$^{21}$ cm$^{-2}$)  &(M$_\odot$yr$^{-1}$)& &\\
\hline
\hline
KK 44 &$-$11.85 &19 & 307 &1.0 &0.6&0.4 &$-$3.44&&9\\
Leo A & $-$11.36&78 & 262 &2.1 & 1.4&0.8 &$-$3.68&7.3&2,8\\
KDG 52 &$-$11.49 &16& 275 &0.6& 0.8&$-$&$>-5.1$&&9\\
UGC 4459 &$-$13.37& 18& 310 &3.2&3.9 &1.6& $-$2.04&7.52&6\\
CGCG 269-049 &$-$12.46&18& 297 &2.4&2.5&1.7&$-$3.08&7.43&3,9\\
UGC 7298 &$-$12.27& 15& 306&1.5&1.9 &0.8&$-$4.5&&9\\
GR 8 &$-$12.11 &30& 305 &1.1 & 1.4&0.9&$-$2.46&7.68&4\\
KK 230 & $-$9.55&33& 304 &0.5 &0.4&$-$&$>-5.53$&&9\\
Sag DIG &$-$11.49 &67 & 325  & 0.6&0.7 & 0.6&$-$3.56& 7.42&1,7\\
DDO 210 &$-$11.09& 61& 291 &1.07&1.3& 1.2 &$-$5.42& 7.4&5,8\\
\hline
\end{tabular}
\end{center}
References:~
1-Hunter \& Elmegreen(2004),
2-James et al.(2004),
3-Kniazev et al.(2003),
4-Legrand et al.(2001),
5-Mateo (1998),
6-Pustilnik et al.(2003),
7-Skillman et al.(1989),
8-Taylor et al.(1998),
9-This paper.
\end{table*}

\begin{figure}
\begin{center}
\psfig{file=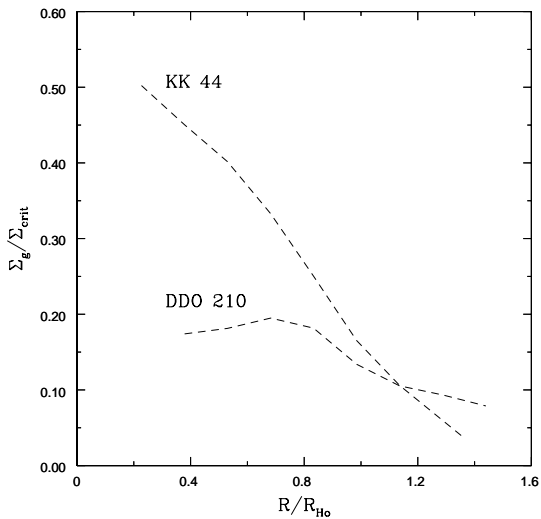,width=3.0truein}
\caption{The ratio between the gas surface density and the Toomre's instability 
threshold density for our sample galaxies DDO 210 and KK 44, which show systematic
rotation. The HI distribution used for deriving the SMD profiles are 
61$^{\prime\prime}\times56^{\prime\prime}$ (DDO 210) and 31$^{\prime\prime}\times
29^{\prime\prime}$ (KK 44).
}
\label{fig:smd_ratio}
\end{center}
\end{figure}

         DDO~210 and KK 44 are the only galaxies in our sample which 
show systematic rotation; the rotation curves of these galaxies have been
presented in  Begum et al.(2003) (KK~44) and Begum \& Chengalur (2004) (DDO 210). 
For these two galaxies we show in Fig.~\ref{fig:smd_ratio}, the ratio of the
azimuthally averaged gas density to the threshold density predicted from 
Toomre's instability criterion. For both of these dwarf galaxies, this ratio
is everywhere smaller than the threshold ratio for star formation in 
spiral galaxies ($\rm{\Sigma_g/\Sigma_{crit} \sim 0.7}$; Kennicutt (1989)).
A similar result was obtained by van Zee et al.(1997), albeit for brighter
(and more rotation dominated, $V/\sigma  > 5$ ) dwarfs. While this
low ratio of gas density to critical density is interesting,
it is unclear whether the Toomre's instability criteria is relevant
in a situation where the rotation speed is comparable
to the velocity dispersion. 

          From the H$\alpha$ overlays, one can see that if there does
exist a threshold HI column density for star formation, it is only in
the very loose sense that one can find a (relatively) high HI column
density contour that just encloses all the star forming regions. The 
actual value of the column density delineating the star forming regions 
varies by more than a factor of 4 between different galaxies in our
sample. Further, the morphologies of the H$\alpha$ emission and the 
high column density HI is quite dissimilar in several cases (e.g. 
UGC~7298, GR 8, KK 44). Thus, while high HI column density may be necessary
for star formation, it clearly is not, in this sample at least, a 
sufficient criteria for star formation. From Table~\ref{tab:result5} 
one can also see that there is no particular correlation between the 
``threshold'' column density and the metallicity. The metallicity of 
our sample galaxies is somewhat lower than that in the original 
sample of Skillman (1987). In that sample the galaxies, with one 
exception (Sextans A, with $12+\log(\rm{O/H})= 7.49$ (Kunth \& O\"stlin (2000)),
have $12+\log(\rm{O/H})$ between $\sim 8$ and  $8.34$. For the galaxies
in our sample for which measurements exist, the metallicity is typically
1~dex lower, while the star formation ``threshold'' density is similar
to that noted by Skillman (1987). It has also been suggested that
the threshold density for star formation is more related to the
presence of a cold phase; in this case, the value of this threshold
does not change much with metallicity (Schaye (2004)).

\begin{figure*}
\begin{center}
\psfig{file=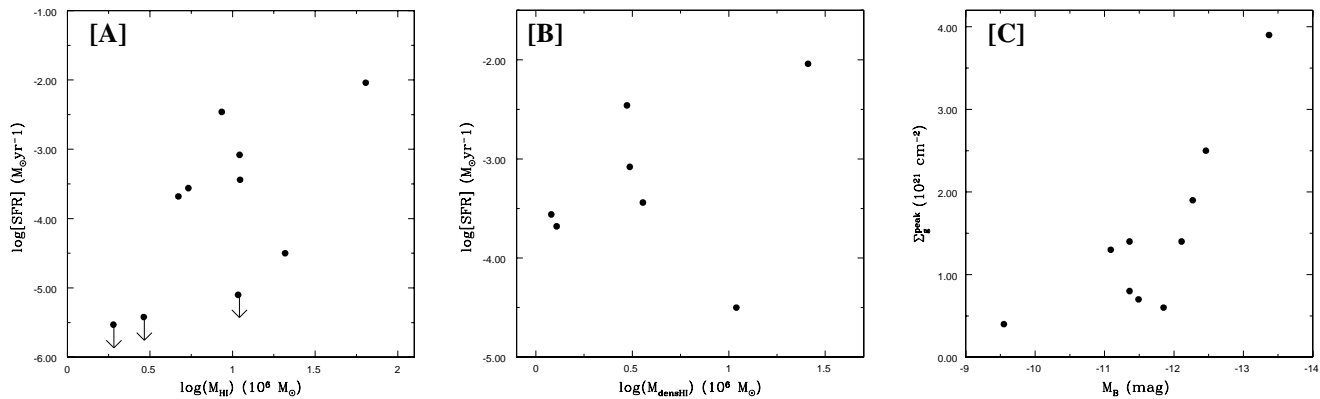,width=7.0truein}
\caption{
{\bf{[A]}} Log of HI mass as a function of SFR for our sample galaxies. In the case of KDG 52, 
KK 230 and DDO 210, the limits on the SFR are shown.
{\bf{[B]}}HI mass of the dense gas i.e mass  of HI gas which has a column density greater 
than the ``threshold'' density, defined in Table~\ref{tab:result5}, as a function of SFR for 
the same sample. {\bf{[C]}} The peak surface gas density as a function of absolute blue magnitude.  
}
\label{fig:corr}
\end{center}
\end{figure*}

     To further explore the connection between the amount of high column density
gas in the galaxy and the star formation, we show in Fig.~\ref{fig:corr}[A] 
and Fig.~\ref{fig:corr}[B], the star formation rate as a function of the
total HI mass as well as the star formation rate as a function of the
mass of HI which has a column density greater than the ``threshold''
density defined in Table~\ref{tab:result5}. The SFR rate actually
correlates slightly better with the total HI mass of the galaxy
(correlation coefficient $\sim 0.34$, excluding those galaxies where
no H$\alpha$ emission was detected) as compared to the mass of the
gas at high HI column density (correlation coefficient $\sim 0.25$).
Clearly, the efficiency with which gas is converted into stars
in these dwarf galaxies is not a function of the amount of high
column density of the gas alone. The strongest correlation 
between the gas distribution and indicators of current or past 
star formation that we find in our sample is that between the 
peak gas density (recall that this need not occur at the center
of the galaxy) and the absolute magnitude (Fig.~\ref{fig:corr}[C]).  
The reason for the existence of such a correlation is unclear. The 
most straight forward interpretation is that bigger galaxies 
are more able to support high column density gas; this in turn
made them, on the average, more efficient at converting their 
gas into stars. On the other hand, as noted above, the current 
star formation rate itself does not correlate particularly strongly 
with the local column  density. One way to reconcile this would
be if in such small galaxies, feedback processes rapidly destroy 
the correlation between the local gas column density and the 
local star formation rate.

\begin{figure*}
\psfig{file=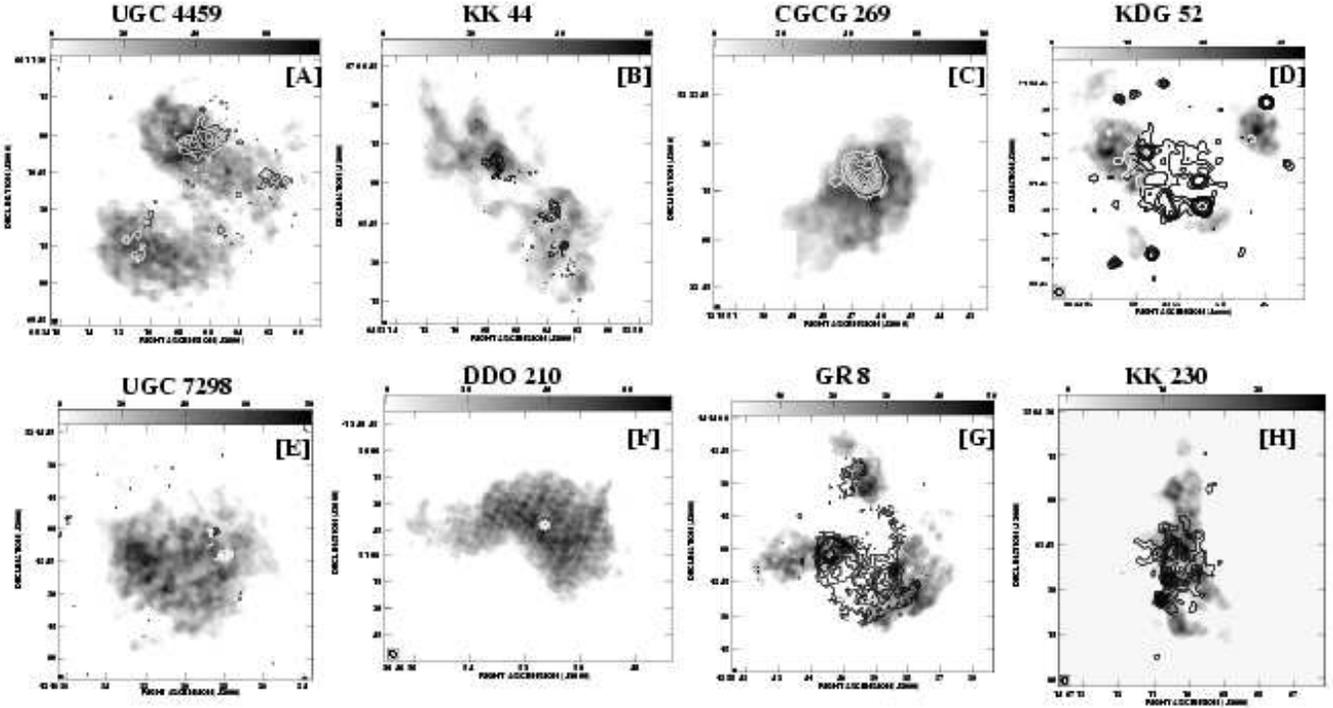,width=7.0truein}
\caption{ The GMRT integrated high resolution HI images of our sample galaxies (greyscales) overlayed 
on  H$\alpha$ images (contours). In case of KK 230 and KDG 52, as no H$\alpha$ emission was detected, 
contours represent the optical broad band emission.  The angular resolutions of the GMRT HI images  
are 3$^{\prime\prime} 
\times 3^{\prime\prime}$ (UGC 4459), 4$^{\prime\prime} \times 4^{\prime\prime}$ (KK 44), 
4$^{\prime\prime} \times 3^{\prime\prime}$ (CCG 269-049), 6$^{\prime\prime} \times 
6^{\prime\prime}$ (KDG 52), 4$^{\prime\prime} \times 4^{\prime\prime}$ (UGC 7298),
4$^{\prime\prime} \times 3^{\prime\prime}$ (DDO 210), 4$^{\prime\prime} \times 
3^{\prime\prime}$ (GR 8) and 4$^{\prime\prime} \times 3^{\prime\prime}$ (KK 230).
The source of H$\alpha$ emission  found in DDO 210 (van Zee 2000) is show as a star.}
\label{fig:overlay_HII}
\end{figure*}

        If feedback from star formation is important, one might expect
to see the strongest evidence for this on small scales. We hence compare
the highest resolution HI images of our sample galaxies with the 
sites of active star formation i.e. H$\alpha$ emission. Our highest resolution
images have beam sizes $\sim 3^{\prime\prime} - 4^{\prime\prime}$, which
corresponds to linear scales between 19~pc and 100~pc for our sample
galaxies. At this high resolution, the emission could not be CLEANed; moment
maps were instead made from the dirty cubes. As discussed earlier, this leads
to a scaling uncertainty, which means we cannot translate the observed
flux distributions into corresponding HI column densities. We can however,
still use our maps to search for correspondences between the morphologies
of the H$\alpha$ and the high column density HI. The overlays are shown 
in Fig.~\ref{fig:overlay_HII}; for all the
galaxies in our sample, the HI emission shows substantial fine scale structure,
with shell like, filamentary as well as discrete clump like morphologies being
visible. The few other dwarf galaxies that have been imaged at similar linear
scales, e.g. IC10 (20 pc; Wilcot \& Miller (1998)), SMC (28 pc; Staveley-Smith et al. (1997)),
LMC (15 pc; Kim et al. (2003)) also show a similar wealth of small scale
structure.  At these scales, the H$\alpha$ emission is sometimes seen coincident
with high HI column densities (e.g. the northern star forming region in UGC~4459, 
north-eastern region in KK~44, the H$\alpha$ knot in CGC~269-049), sometimes the
high HI column density gas forms a  shell around the H$\alpha$ emission (e.g.
the south-western star forming region in KK~44, the high density HI clumps in
GR~8) and sometimes there seems to be no connection at all between the high
density HI gas and the H$\alpha$ emission (e.g. UGC~7298). In general, while
high HI column density gas is in general present in the vicinity (as measured
on linear scales $< 100$~pc) of current star formation, there does not seem
to be a simple, universal relationship between the H$\alpha$ emitting gas
and the high column density neutral gas.

\subsection{HI line profiles and star formation}
\label{ssec:starformation}

Young and collaborators have found that in faint dwarfs, HI line profiles in regions 
of active star formation differ substantially from a simple Gaussian shape
( Leo A, Young \& Lo 1996; Sag~DIG, Young \& Lo 1997; UGCA~292 Young et al. 2003). 
Young et al.(2003) used both a two Gaussian fit as well as a fit using Gauss-Hermite 
polynomials to parametrize the line profiles. To facilitate easy
comparison, we fit the line profiles to our sample galaxies using the same
two models. Apart from Leo~A and Sag~DIG (where we use essentially the same
VLA data as used by Young et al.) our sample has two galaxies, viz. DDO~210 and
GR~8, in common with the earlier sample. We include these two galaxies in
the analysis that we do in this section, and compare the results obtained from
the GMRT data with those obtained earlier with the VLA.

   Ideally one would like
to fit profiles to data cubes which correspond to the same linear resolution
at the distance to the galaxy. However  fitting to the line profile requires 
a good signal to noise ratio, and  for the fainter galaxies, the signal to
noise ratio  is adequate only in the lowest resolution images. The linear 
resolution of the data cubes used for profile fitting is given in Table~\ref{tab:result6}. 
For all the sample galaxies, the line profiles at each location were 
first fitted with a single Gaussian component and the residuals were 
inspected. HI profiles in some cases were found to deviate measurably
from a simple Gaussian $-$ in such cases the profiles were often either 
asymmetric or symmetric, but with narrower peak and broader wings than
a Gaussian.

       The profiles were then fit with a double Gaussian and also separately with a Gauss-Hermite 
polynomial. The Gauss-Hermite polynomial used for the profile fitting is given
as:

\begin{equation}
\phi(x)=ae^{\frac{-1}{2}y^2}[1+\frac{h_3}{\sqrt 6}(2 \sqrt2 y^3-3 \sqrt2 y)+
           \frac{h_4}{\sqrt24}(4y^4-12y^2+3)]
\label{eqn:hermite}
\end{equation}

where y$\rm{=\frac{(x-b)}{c}}$. Parameters a, b, c are equal to the amplitude, mean and 
dispersion respectively for a Gaussian (to which the Eqn.(~\ref{eqn:hermite}) reduces to,
when parameters $h_3$ and $h_4$ are zero). Parameter $h_3$ is related to the skewness 
of the line profile i.e. in  the case of an asymmetric line profile $h_3\neq0$. If 
$h_4\neq$0, the line profiles either have a more pointed top with broader wings 
($h_4>0$) or have a flatter top  ($h_4<0$) than a Gaussian. 

In the case of Gauss-Hermite fits, the profiles for which both $h_3$ and $h_4$ 
parameters were less than 3 times the uncertainty in these parameters, were rejected as 
bad fits. Similarly, in the case of double Gaussian fit, the profiles for which the width 
of the fitted narrow component was less than the velocity resolution of 1.65 \kms (within 
the errorbars)  were rejected.  Following Young \& Lo (1996), we use the F-test to distinguish 
between  profiles that are adequately fit by a single Gaussian and those which are not.
Locations where the null hypothesis (viz. that a single Gaussian provides an equally 
good description of the line profile as compared to a double Gaussian or Gauss-Hermite 
polynomial) was rejected at the 90\% or higher confidence level were compared with the 
locations of on-going star formation.

\begin{figure*}
\psfig{file=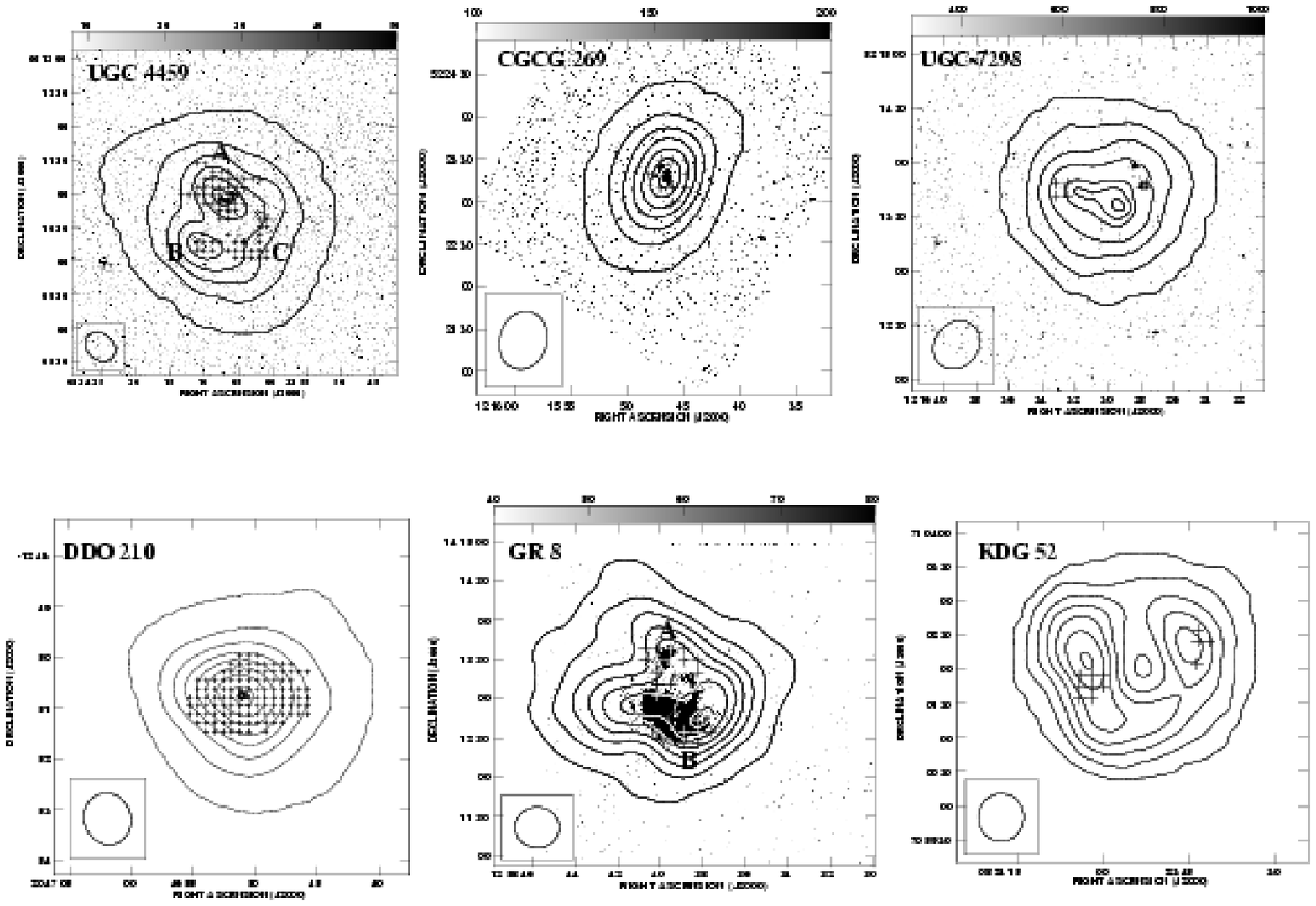,width=7.0truein}
\caption{ GMRT integrated HI column density distribution for our sample galaxies (contours) 
overlayed on the H$\alpha$ emission (greyscale) from the galaxies. The regions where the 
line profile deviated from the single Gaussian are marked as crosses. In the case 
of DDO 210, the HII region the galaxy is marked as stars. In case of KDG 52,
no H$\alpha$ emission was detected in the galaxy. The resolution of the HI distribution 
are 29$^{\prime\prime} \times 27^{\prime\prime}$
(UGC 4459),  42$^{\prime\prime} \times 39^{\prime\prime}$ (CGCG 269-049),  28$^{\prime\prime} \times
24^{\prime\prime}$ (UGC 7298),  60$^{\prime\prime} \times 58^{\prime\prime}$ (DDO 210),  
30$^{\prime\prime} \times 30^{\prime\prime}$ (GR 8) and  42$^{\prime\prime} \times 
39^{\prime\prime}$ (KDG 52).}
\label{fig:gauss}
\end{figure*}

The results of the line profile fitting are given in Table~\ref{tab:result6}. 
Col.(1) gives the galaxy name, Col.(2) shows the resolution of the HI distribution 
used for the profile fitting, Col.(3) the liner resolution in pc, Col.(4) the minimum 
gas surface density (the observed HI column density corrected for inclination and He 
content) enclosing the regions with non Gaussian HI profiles, Col.(5) $h_3$ and $h_4$ 
parameters of the best fit Gauss-Hermite polynomial, Col.(6) range of the velocity 
dispersion of the narrow component in the double Gauss fit and Col.(7) range of the 
velocity dispersion of the broad component in the double Gauss fit. In case of UGC~4459 
and GR~8, results of the line profile fitting from separate regions (as marked 
in the Fig.~\ref{fig:gauss}), are described separately in the Table~\ref{tab:result6}. 
For KDG~52, the two separate regions showing deviation from the Gaussian, near the 
eastern and western clump, gave similar results.

\begin{table*}
\begin{center}
\caption{Results of the profile fitting to our sample galaxies}
\label{tab:result6}
\vskip 0.1in
\begin{tabular}{|lcccccc|}
\hline
Galaxy& Beam & Linear resolution &$\rm{\Sigma_{gas}^c}$ &Gauss-Hermite fit & Narrow component & Broad component \\  
& (arcsec) & (pc)& ($10^{21}$ cm$^{-2}$) &&(\kms)&(\kms)\\
\hline
\hline
KK 44 & 40  & 647&$-$ &$-$&$-$&$-$\\
KDG 52 & 42& 723&0.2&$h_3=0~~h_4>0$&2.0$-$4.5&8$-$14\\
UGC 4459 (A) & 26& 450&1.0 &$h_3<0~~h_4=0$&3.5$-$7.0&6$-$17\\
UGC 4459 (B) & 26& 450&2.0 & $h_3=0~~h_4>0$&6.0&10$-$17\\
UGC 4459 (C) & 26& 450&1.0 & $h_3=0~~h_4>0$&4.5$-$6.0&10$-$16\\
CGCG 269-049 & 42& 692& 0.4&$h_3<0~~h_4=0$&2$-$5&7$-$13\\
UGC 7298 & 26& 529&1.0 &$h_3=0~~h_4>0$&2.5$-$4 &9.5$-$11\\
GR 8 (A) & 30& 305&0.7 &$h_3<0~~h_4>0$ &2$-$6&7$-13$\\
GR 8 (B) & 30& 305&0.9 &$h_3=0~~h_4>0$ &2$-$3&8$-9.5$\\
KK 230 & 48& 489&$-$ &$-$&$-$&$-$\\
DDO 210 & 60& 291& 0.4 & $h_3=0~~h_4>0$&3$-$6&8$-$14\\
\hline
\end{tabular}
\end{center}
\end{table*}

The regions in our sample galaxies where the double Gaussian gave a better fit to the
line profiles than a single Gaussian are marked as crosses on the HI column density 
distribution  in Fig.~\ref{fig:gauss}. The regions where the Gauss-Hermite polynomial 
gave a better fit than the single Gaussian are almost similar to the regions where 
the double Gaussian gave a good fit, hence are not shown separately. We note that
for most galaxies (with the exception of DDO~210) the extent of these regions is
comparable to our spatial resolution. To allow easy cross comparison with regions
having on-going star formation, the H$\alpha$  emitting regions are represented as 
greyscales in Fig.~\ref{fig:gauss}.  The line profiles for KK~44 and KK~230 
throughout the galaxy are found to be well described by a single Gaussian component, 
hence are not shown. 

As seen in Fig.~\ref{fig:gauss}, no particular correlation is seen between the 
location of H$\alpha$ emission and the deviation of HI line profiles from single Gaussians. 
Not all star forming regions in our sample galaxies show deviation of the line 
profiles e.g. UGC~7298, KK~44 and eastern clump in GR~8. Conversely, not all regions 
which  show deviations of line profiles are associated with the  star forming regions 
e.g. KDG~52, DDO~210  and UGC~7298. In this sample at least, the correlation found
by Young et al.(2003) from their analysis of 3 dwarf irregular galaxies does not
seem to hold.  We also do not find any correlation between the presence of
asymmetric profiles and the global star formation activity. 

       The dwarf galaxies DDO~210 and GR~8 are common between our sample and 
that of Young et al.(2003). The results derived by Young et al.(2003) for DDO~210 
are similar to our results. On the other hand, for GR~8, Young et al.(2003) found few 
line profiles with  $h_3\neq 0$ in the southern clump, (albeit with a very small 
magnitude of $h_3$), and almost none in the eastern clump, while we  found all the 
profiles associated with the southern and eastern clump have $h_3=0$, (within  the 3$\sigma$ 
uncertainty of the parameter). However, this difference is not pronounced, and
one should also note that Young et al.(2003) used 14$^{\prime\prime} \times 14^{\prime \prime}$ 
and 18$^{\prime\prime} \times 18^{\prime \prime}$ resolution data cubes, whereas we have 
used a 30$^{\prime\prime} \times 30^{\prime \prime}$ resolution data cube. In general
hence, there is relatively good agreement between the fits obtained with the
GMRT and VLA data.

\section{Summary}

       We compare the HI distribution, kinematics and current star formation in
a sample of 10 extremely faint nearby dwarf galaxies. For 5 of these galaxies, fresh
GMRT HI data are presented in this paper. The large scale gas distribution in the
galaxies is generally clumpy, and the peak HI column density is often well removed from
the geometric center. For all galaxies we find a large scale ordered velocity field,
although the patterns are mostly not reconcilable with that expected from a rotating 
disk. From a simplistic virial theorem based estimate of the dynamical mass, we find 
very tentative evidence that the faintest dwarf irregulars have a somewhat smaller 
baryon fraction than brighter galaxies.We compare the regions of ongoing star formation
with regions of high HI column density, with the column density being measured
at a uniform linear scale ($\sim 300$~pc) for all galaxies in our sample. We find
that while the H$\alpha$ emission is confined to regions with relatively high
column density, in general the morphology of the H$\alpha$ emission is not 
correlated with that of the high column density HI gas. Thus, while high
gas column density may be a necessary condition for star formation, it is
not, in this sample at least, a sufficient condition. We also examine the line
profiles of the HI emission, and check if deviations from a simple Gaussian
profile is correlated with star formation activity. We do not find any such
correlation in our sample -- there are regions with on-going star formation
but with simple Gaussian line profiles, as well as regions with complex
line profiles but no ongoing star formation. Finally, we look at the distribution
of HI gas at linear scales $\sim 20-100$ pc. All our sample galaxies show
substantial small scale structures with shell like, filamentary as
well as clumpy features being identifiable in the images. H$\alpha$ emitting
regions are sometimes associated with clumpy features; sometimes the
H$\alpha$ emission lies inside a shell like feature in the HI, and sometimes
there is no particular HI column density enhancement seen near the H$\alpha$
emission. The interplay between star formation and gas density and kinematics
in these galaxies hence appears to be very varied, and the general unifying
patterns seen in larger irregulars and spiral galaxies seem to be absent.
Star formation and feedback are complex processes, and perhaps it is
the presence of simple large scale correlations in big galaxies that should
surprise us more than the absence of such correlations in small galaxies.

\section*{Acknowledgments}
We would like to thank Dr U. Hopp for providing the optical 
images of UGC 4459 and Dr S. Pustilnik for useful discussion on 
UGC 4459. AB thanks the Kanwal Reiki Scholarship of TIFR
for partial financial support. 
        The observations presented in this paper would not have been
possible without the many years of dedicated effort put in by the
GMRT staff in order to build the telescope. The GMRT is operated
by the National Center for Radio Astrophysics of the Tata Institute
of Fundamental Research.

\end{document}